\documentclass{ws-ijmpcs}

\begin{document}

\markboth{Marin-Slobodan Toma\v s} {Casimir effect across a
layered medium}

\catchline{}{}{}{}{}
%
%%%%%%%%%%%%%%%%%%%%%%%%%%%%%%%%%%%%%%%%%%%%%%%%%%%%%%%%%%%%%%%%%%%%

\title{Casimir effect across a layered medium}
 \author{Marin-Slobodan Toma\v s}

 \address{Rudjer Bo\v skovi\' c Institute, P. O. B. 180,
 10002 Zagreb, Croatia\\tomas@thphys.irb.hr}

\maketitle

%\begin{history}
%\received{Day Month Year} \revised{Day Month Year}
%\end{history}

\begin{abstract}
Using nonstandard recursion relations for Fresnel coefficients
involving successive stacks of layers, we extend the Lifshitz
formula to configurations with an inhomogeneous, n-layered, medium
separating two planar objects. The force on each object is the sum
of a Lifshitz like force and a force arising from the
inhomogeneity of the medium. The theory correctly reproduces very
recently obtained results for the Casimir force/energy in some
simple systems of this kind. As a by product, we obtain a formula
for the force on an (unspecified) stack of layers between two
planar objects which generalizes our previous result for the force
on a slab in a planar cavity.

\end{abstract}

\ccode{PACS numbers: 12.20.Ds, 42.25.Bs, 42.25.Gy}

\section{Introduction}

Very recently, several papers appeared dealing with the theory of
the Casimir effect in systems consisting of two perfectly
reflecting plates separated by a layered
medium\cite{Teo,Kheir,Amoog}. Among the other results, these works
provided formulas for the Casimir force and/or energy for a few
simple systems of this sort (with up to five layers
media\cite{Teo} between the plates). Using the theory of the
Casimir effect in multilayers\cite{Tom02} and nonstandard
recursion relations for Fresnel
coefficients\cite{Tom02,Tom95,Kat,Tom10}, in this Note we derive
formulas for the Casimir force and energy for systems with
arbitrary plates separated by arbitrary inhomogeneous, generally
$n$-layered, media.
\begin{figure}[htb]
 \begin{center}
 \resizebox{7cm}{!}{\includegraphics{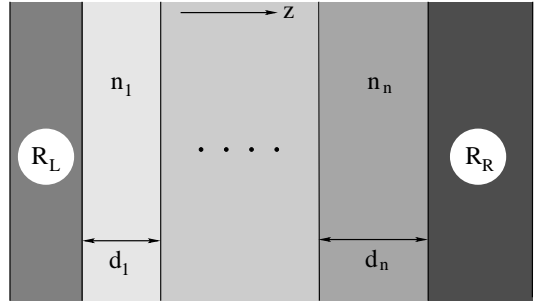}}
 \end{center}
 \caption{\label{sys}Two plates separated by a layered medium shown
 schematically. Plates are described by their reflection coefficients
 $R_L$ and $R_R$ and the medium layers by their (complex) refraction
 indexes $n_a(\omega)=\sqrt{\varepsilon_a(\omega)\mu_a(\omega)}$,
 $a=1\dots n$.}
\end{figure}

\section{Casimir effect across a layered medium}
Consider the system consisting of two planar objects (plates)
separated by a layered medium, as depicted in Fig. \ref{sys}.
According to the theory of the Casimir effect in
multilayers\cite{Tom02}, the Casimir forces on the left ($L$) and
the right ($R$) plate are given by
\begin{equation}
\label{FLR} F_L\equiv F_{1-}=T^{(1)}_{zz}\hspace{0.5cm} {\rm
and}\hspace{0.5cm} F_R\equiv F_{n+}=-T^{(n)}_{zz},
\end{equation}
respectively, where (unless necessary, we omit the polarization
index $q=p,s$ when writing Fresnel coefficients)
\begin{equation}
\label{Tj} T^{(j)}_{zz}=\frac{\hbar}{2\pi^2}\int_0^\infty d\xi
\int^\infty_0 dkk\kappa_j\sum_{q} \frac{r_{j-}r_{j+}e^{-2\kappa_j
d_j}} {1-r_{j-}r_{j+}e^{-2\kappa_j d_j}}
\end{equation}
is the relevant component of the vacuum-field (Minkowski) stress
tensor in the layer $j$. Here
$\kappa_j=\sqrt{n^2_j(i\xi)\xi^2/c^2+k^2}$ is the perpendicular
wave vector at the imaginary frequency ($\omega=i\xi$) in the
layer, $k=\sqrt{k^2_x+k^2_y}$ is the magnitude of the wave vector
parallel to the system surfaces and $r_{j\pm}(i\xi,k)$ are the
reflection coefficients of the right and left stack of layers
bounding the layer $j$. These reflection coefficients obey
generalized recursion relations \cite{Tom02,Tom95,Kat,Tom10}
\begin{equation}
\label{RR} r_{j\pm}=r_{j/l}+\frac{t_{j/l}t_{l/j}r_{l\pm}
 e^{-2\kappa_ld_l}}{1-r_{l/j}r_{l\pm}e^{-2\kappa_ld_l}},\;\;\;
 %\label{rj-l-m}
t_{j\pm}=\frac{t_{j/l}t_{l\pm}e^{-\kappa_ld_l}}
{1-r_{l/j}r_{l\pm}e^{-2\kappa_ld_l}},
%\label{tj-l-m}
\end{equation}
where $l$ denotes an intermediate layer and where the symbol
$a/b\equiv a\ldots b$ is used to denote the stack of layers
between layers $a$ and $b$. As seen, these recurrence relations
look the same as the standard ones \cite{BW,Chew} (to which they
reduce in case that layers $j$ and $l$ are neighbor layers),
however, this time they generally involve Fresnel coefficients
$r_{j/l}$, $r_{l/j}$, $t_{j/l}$ and $t_{l/j}$ of the {\it stack}
between the layers $j$ and $l$.

Using the above recursion relations,  reflection coefficients
$r_{1+}$ and $r_{n-}$ can be expressed as
\begin{equation}
\label{rr}
 r_{1+}=\frac{r_{1/n}+a_{1/n}
 R_Re^{-2\kappa_nd_n}}
 {1-r_{n/1}R_Re^{-2\kappa_nd_n}},\;\;\;
r_{n-}=\frac{r_{n/1}+a_{n/1}
 R_Le^{-2\kappa_1d_1}}
 {1-r_{1/n}R_Le^{-2\kappa_1d_1}},
 \end{equation}
where we have introduced the quantity
\begin{equation}
 a_{1/n}=t_{1/n}t_{n/1}-r_{1/n}r_{n/1}=a_{n/1}
\end{equation}
and identified $r_{n+}$ and $r_{1-}$ as the reflection
coefficients $R_R$ and $R_L$, respectively, of the plates.
Therefore, from (\ref{FLR}) and (\ref{Tj}) we obtain for the
forces on the plates
\begin{subequations}
\label{FLR2}
\begin{eqnarray}
 F_L&=&\frac{\hbar}{2\pi^2}\int_0^\infty d\xi
\int^\infty_0 dkk\kappa_1\sum_{q} \frac{1}{N_n}
(r_{1/n}+a_{1/n}R_Re^{-2\kappa_n d_n})R_Le^{-2\kappa_1 d_1}, \\
F_R&=&\frac{-\hbar}{2\pi^2}\int_0^\infty d\xi \int^\infty_0
dkk\kappa_n\sum_{q} \frac{1}{N_n}
(r_{n/1}+a_{1/n}R_Le^{-2\kappa_1d_1})R_Re^{-2\kappa_n d_n},
\end{eqnarray}
\end{subequations}
where
\begin{equation}
\label{Nn}
N_n=1-(r_{1/n}R_Le^{-2\kappa_1d_1}+r_{n/1}R_Re^{-2\kappa_n d_n})-
a_{1/n}R_LR_Re^{-2\kappa_1 d_1-2\kappa_n d_n}.
\end{equation}
As seen, since $r_{1/n}\neq r_{n/1}$ unless the medium is not
symmetric across the gap, these forces are not generally equal in
magnitude and each of them consists of a Lifshitz-like force
(given by the second terms in (\ref{FLR2})) and a force due to the
inhomogeneity of the medium. We also note that owing to the medium
inhomogeneity there is a force\cite{Tom02}
$F_S=T^{(n)}_{zz}-T^{(1)}_{zz}=-F_R-F_L$ on the central stack of
the medium given explicitly by
\begin{eqnarray}
F_S&=&\frac{\hbar}{2\pi^2}\int_0^\infty d\xi \int^\infty_0
dkk\sum_{q} \frac{1}{N_n} (\kappa_nr_{n/1}R_Re^{-2\kappa_n
d_n}-\kappa_1r_{1/n}R_Le^{-2\kappa_1 d_1})\nonumber\\
&&+(\kappa_n-\kappa_1)\frac{a_{1/n}}{N_n}R_LR_Re^{-2\kappa_1d_1
-2\kappa_nd_n}.
\end{eqnarray}
When $n_1=n_n$, this generalizes previously obtained result for
the force on a slab in a planar cavity\cite{Tom02,Tom10,Elli} to
configurations with the slab replaced by an unspecified
multilayered stack.

%\section{Casimir energy}
Having determined forces on the plates, we can calculate the
Casimir energy of the system from\cite{Tom02} $F_L=\partial
E/\partial d_1$ or $F_R=-\partial E/\partial d_n$. We obtain
\begin{equation}
\label{E}
E=\frac{\hbar}{(2\pi)^2}\int_0^\infty d\xi \int^\infty_0
dkk\sum_q \ln N_n.
\end{equation}
In the following, we illustrate this formula by discussing its
implications for several simple systems and comparing them with
the results obtained in Refs. \refcite{Teo}-\refcite{Amoog}.

\section{Discussion}
It is easy to see that (\ref{Nn}) and (\ref{E}) give correctly the
Casimir energy in case of a homogeneous medium between the plates.
Indeed, assuming that all $n$ medium layers are made of the same
matter ($n_a=n$, $\kappa_a=\kappa$), we have $r_{1/n}=r_{n/1}=0$
and $t_{1/n}=t_{n/1}=\exp{[-\kappa (d-d_1-d_n)]}$ so that
$a_{1/n}=\exp{[-2\kappa (d-d_1-d_n)]}$ and
\begin{equation}
N_n=1-R_LR_Re^{-2\kappa d},
\end{equation}
where $d$ is the distance between the plates. This leads to the
standard Lifshitz-type formula\cite{Dzy} for the Casimir energy.
For perfectly reflecting plates\cite{Kheir,Amoog}, we must let
here $R_LR_R=1$.

In the $n=2$ case (two media between the plates), we have
$r_{1/2}=r_{12}=-r_{21}=-r_{2/1}$ and
$t_{1/2}=t_{12}=(\mu_2\kappa_1/\mu_1\kappa_2)t_{21}
=(\mu_2\kappa_1/\mu_1\kappa_2)t_{2/1}$, where $r_{12}$ and
$t_{12}$ are the {\it single-interface} Fresnel
coefficients\cite{Tom02,Tom95}
\begin{equation}
 \label{sic}
 r_{12}=\frac{\kappa_1-\gamma_{12}\kappa_2}
 {\kappa_1+\gamma_{12}\kappa_2}=-r_{21},\;\;\;
 t_{12}=\sqrt{\frac{\gamma_{12}}{\gamma^s_{12}}}(1+r_{12})=
 \frac{\mu_2\kappa_1}{\mu_1\kappa_2}t_{21},
 \end{equation}
 with $\gamma^p_{12}=\varepsilon_1/\varepsilon_2$ and
 $\gamma^s_{12}=\mu_1/\mu_2$. Noting that $a_{1/2}=1$,
 we have
%\begin{equation}
%F_{L(R)}=\frac{\hbar}{2\pi^2}\int_0^\infty d\xi \int^\infty_0
%dkk\kappa_{1(2)}\sum_{q} \frac{1}{N_2}
%(r_{12}R_Le^{-2\kappa_{1(2)} d_{1(2)}}\pm
%R_LR_Re^{-2(\kappa_1d_1+\kappa_2 d_2)}),
%\end{equation} where
\begin{equation}
\label{N2} N_2=1-r_{12}(R_Le^{-2\kappa_1 d_1}-R_Re^{-2\kappa_2
d_2}) -R_LR_Re^{-2(\kappa_1 d_1+\kappa_2 d_2)},
\end{equation}
which, in conjunction with (\ref{E}), gives the Casimir energy for
the present system. This result coincides with the corresponding
result obtained in Ref. \refcite{Kheir} providing that we let for
perfectly reflecting plates $R^q_{L(R)}=-1$. We note, however,
that perfect reflectors are standardly simulated by media with
infinitely large permittivities (conductivities) in which case
(\ref{sic}) implies ($\varepsilon_2\rightarrow\infty$) that
$R^s_{L(R)}=-1$ but $R^p_{L(R)}=1$. Therefore, with this
convention, our result disagrees with that of Ref. \refcite{Kheir}
regarding the $p$ contribution to the Casimir force/energy.

Using recursion relations (\ref{RR}), E for more complex ($n\geq
3$) systems can be written in terms of lower-layered stacks and,
owing to the number of the medium layers, this can be done in a
number of ways. Clearly, to obtain the effective Casimir energy,
we can drop from these results the terms not involving $d_1$ or
$d_n$. Thus, for example, from (\ref{RR}) we have\cite{Tom10}
\begin{subequations}
%\label{r123}
\begin{eqnarray}
 r_{1/n}&=&\frac{r_{1/l}+a_{1/l}r_{l/n}e^{-2\kappa_ld_l}}
 {D_l},\hspace{1cm}
r_{n/1}=\frac{r_{n/l}+a_{n/l}r_{l/1}e^{-2\kappa_ld_l}}
 {D_l},\\
 a_{1/n}&=&\frac{a_{1/l}a_{n/l}e^{-2\kappa_ld_l}-r_{1/l}r_{n/l}}
 {D_l},\hspace{1cm}D_l=1-r_{l/1}r_{l/n}e^{-2\kappa_ld_l}.
 \end{eqnarray}
 \end{subequations}
Using this in (\ref{Nn}) and rearranging, we find
\begin{equation}
N_n=\frac{N^{(l)}_n}{D_l},
\end{equation}
where
\begin{eqnarray}
\label{Nkn} N^{(l)}_n&=&(1-r_{1/l}R_Le^{-2\kappa_1
d_1})(1-r_{n/l}R_Re^{-2\kappa_n d_n})\nonumber
\\
&-&e^{-2\kappa_l d_l}(a_{1/l}R_Le^{-2\kappa_1
d_1}+r_{l/1})(a_{n/l}R_Re^{-2\kappa_n d_n}+r_{l/n}).
\end{eqnarray}
Finally, inserting this $N_n$ in (\ref{E}) and dropping the
(ineffective) term involving $D_l$, we find for the effective
Casimir energy $E_l$ of the system (with respect to the layer $l$)
\begin{equation}
\label{Ek} E_l=\frac{\hbar}{(2\pi)^2}\int_0^\infty d\xi
\int^\infty_0 dkk\sum_q \ln N^{(l)}_n.
\end{equation}
This generalizes the ($T=0$) result for the Casimir interaction
energy between two slabs obtained in Ref \refcite{Teo} using a
realistic Casimir piston approach and a five layer model for the
medium to arbitrary multilayered slabs and plates. Note that, when
removing the plates by letting $d_{1(n)}\rightarrow\infty$, we
have $N^{(l)}_n\rightarrow D_l$ and (\ref{Nkn}) and (\ref{Ek})
give the Casimir interaction energy of the two stacks of layers
separated by a layer of medium $l$, as derived in Ref.
\refcite{Tom02}.

We illustrate the above result by considering the n=3 system. In
this case, there is only one intermediate layer and the effective
Casimir energy (\ref{Ek}) is entirely expressed in terms of the
single-interface reflection coefficients $r_{12}=-r_{21}$ and
$r_{32}=-r_{23}$. From (\ref{Nkn}), we have ($a_{1/2}=a_{3/2}=1$)
\begin{eqnarray}
\label{N23} N^{(2)}_3&=&(1-r_{12}R_Le^{-2\kappa_1
d_1})(1-r_{32}R_Re^{-2\kappa_3 d_3})\nonumber
\\
&-&e^{-2\kappa_2 d_2}(R_Le^{-2\kappa_1
d_1}-r_{12})(R_Re^{-2\kappa_3 d_3}-r_{32}).
\end{eqnarray}
As mentioned, for perfectly reflecting plates we must let here
$R^q_{L(R)}=\delta_{qp}-\delta_{qs}$. This result then coincides
with the corresponding result derived in Ref. 1 whereas the
results obtained in Refs. 2 and 3 correspond to plates with
$R^q_{L(R)}=-1$ and $R^q_{L(R)}=1$, respectively.

\section{Summary}
Effectively, in this work we have extended the Lifshitz formula to
configurations with an inhomogeneous, n-layered, medium separating
two planar objects. The force on each object is the sum of a
Lifshitz-like force and a force arising from the inhomogeneity of
the medium. Owing to this inhomogeneity, there is also a force
acting on the medium. When the first and the last medium layer are
made of the same matter, this result generalizes previously
obtained one for the force on a slab in a planar cavity to
arbitrary multilayered slabs.

\section*{Acknowledgments}

This work was supported by the Ministry of Science, Education and
Sport of the Republic of Croatia under Contract No.
098-0352828-3118.

%\section{References}

 \end{document}